\documentclass[aps,prd,preprintnumbers,nofootinbib,twocolumn]{revtex4}
\usepackage{bm}
\usepackage{latexsym}
\usepackage{dcolumn}
\usepackage{amsmath,amsfonts,amssymb}
\usepackage{graphicx,epsfig}
\usepackage{color}
\usepackage{amsthm}

\def\be {\begin{equation}}
\def\ee {\end{equation}}
\def\bea {\begin{eqnarray}}
\def\eea {\end{eqnarray}}
\def\bc {\begin{center}}
\def\ec {\end{center}}
\def\bfg {\begin{figure}}
\def\efg {\end{figure}}
\def\bi {\begin{itemize}}
\def\ei {\end{itemize}}
\def\nn {\nonumber}

\def\la {\label}
\def\le {\left}
\def\ri {\right}

\def\no {\noindent}

\def\a  {\alpha}

\def\D  {\Delta}

\def\beq{\begin{equation}}
\def\eeq{\end{equation}}
\def\br{\begin{eqnarray}}
\def\er{\end{eqnarray}}
\newcommand{\eel}[1] {\label{#1}\end{equation}}

\newcommand{\bdm}{\begin{displaymath}}
\newcommand{\edm}{\end{displaymath}}

\begin{document}
\title{Emergence of Cosmic Space and Minimal Length in Quantum Gravity}

\author{Ahmed Farag Ali} \email[]{ahmed.ali@fsc.bu.edu.eg; afarag@zewailcity.edu.eg}

\affiliation{Center for Fundamental Physics, Zewail City of Science and Technology, Giza, 12588, Egypt.\\}
\affiliation{Department of Physics, Faculty of Sciences, Benha University, Benha, 13518, Egypt.\\}

\begin{abstract}
\par\noindent
An emergence of cosmic space has been suggested by Padmanabhan in \cite{Padmanabhan:2012ik}.
This new interesting approach argues that the expansion of the universe is due to the difference between
the number of degrees of freedom on a holographic surface and the one in the emerged
bulk. In this paper, we derive, using emergence of cosmic space framework, the general dynamical equation
of FRW universe filled with a perfect fluid by considering a generic form of the entropy as a function of area.
Our derivation is considered as a generalization of emergence of cosmic space with a general form of entropy.
We apply our equation with higher dimensional spacetime and derive
modified Friedman equation in Gauss-Bonnet gravity.
We then apply our derived equation with the  corrected entropy-area law that follows from Generalized Uncertainty Principle (GUP)
and derive a modified Friedmann equations due to the GUP. We then derive the modified Raychaudhuri equation due to GUP in
emergence of cosmic space framework and investigate it using fixed point method.
Studying this modified Raychaudhuri equation leads to nonsingular solutions
which may resolve singularities in FRW universe .
\end{abstract}

\maketitle


\section{Introduction}

Various approaches to understanding the origin of gravity suggest that
gravity is an emergent phenomenon. Most of these approaches depend on the
genuine connection between gravity and the first law of thermodynamics.
This was first realized by Hawking in \cite{Hawking} where
it was proposed that black hole behaves like a black body radiator with a temperature
proportional to its surface gravity and with entropy proportional to the horizon area of
the black hole \cite{Bekenstein}. Motivated by the elegant relation between
entropy and horizon area, Jacobson \cite{Jacobson} found that the
Einstein field equations can follow exactly from the fundamental relation of first
law of thermodynamics which connects heat, entropy, and temperature $dQ=T dS$.
Inspired by Jacobson approach, it was straightforward to derive Friedmann
equations of Friedmann-Robertson-Walker (FRW) universe, from the Clausius
relation with the apparent horizon of FRW universe with assumption that the
entropy is proportional to the area of the apparent horizon \cite{Cai}. Recently, Verlinde \cite{everlind1},
with more rigorous approach, suggested that gravity is not a fundamental force and can be explained as
an entropic force, and he derived Newton's law of gravitation and Einstein equations based on his proposal.
All the mentioned studies are dealing with gravity as an emergent phenomena but they did not touch the spacetime.
It is known from general relativity that spacetime and gravity are quite related.
A naturally arising question about the nature of spacetime should be taken into consideration;
If Gravity is an emergent phenomenon, what about spacetime? Recently, Padmanabhan \cite{Padmanabhan:2012ik}
proposed that cosmic space is an emergent phenomenon with the cosmic time evolution.
His idea was motivated by the spacial role of  cosmic time of a geodesic observer to which the
observed cosmic microwave background (CMB) radiation is homogeneous and isotropic. Based on this, the
expansion of the universe can be realized as a result of the difference between surface degrees of freedom ($N_{\text{sur}}$)
and bulk degrees of freedom ($N_{\text{bulk}}$) in a region of emerged space and using this argument,
the dynamical equation of an FRW universe has been derived successfully.

The scope of the present paper is to derive the general dynamical equation of
FRW universe filled with perfect fluid. This can be done by considering a generic
form of entropy as a function of area and this can
host every possible corrections to the entropy-area law. This derivation will be done through
the framework of emergence of cosmic space that has been proposed by Padmanabhan \cite{Padmanabhan:2012ik}.
We then apply our equation with the corrected entropy which follows from
the generalized uncertainty principle (GUP) \cite{guppapers,advplb,Das:2010zf,kmm,kempf,brau},
where GUP introduces a possible existence of minimal length
which represents a natural cutoff, and it is expected to introduce
a possible resolution to the known curvature singularities in general relativity.

An outline of this paper is as follows. In Section (\ref{cosmic space}), we review the proposal by
\cite{Padmanabhan:2012ik} and introduce the calculations of the difference between surface degrees of freedom and bulk degrees
of freedom which derive at the end the dynamical Friedmann equation.
In Section (\ref{ModFried}), we derive the modified Friedmann equation with
a general form of the entropy as function of the area using emergence of cosmic space framework,
and derive the corresponding dynamical equation of FRW universe filled with a perfect fluid.
In Section (\ref{GUPentropy}), we review the corrected entropy-area
law due to the GUP, and in Section (\ref{FRW-GUP}) we derive the corrected dynamical equation of FRW universe due to the GUP through
emergence of cosmic space framework and we then derive the corresponding Raychaudhuri equation and study its non-singular solutions
using fixed point method.

\section{Emergence of Cosmic Space}
\label{cosmic space}

In this section, we review briefly the proposal of emergence of cosmic space by Padmanabhan \cite{Padmanabhan:2012ik}.
It has been studied a pure de Sitter universe and realized that the holographic principle can take the following form
\bea
N_{\text{sur}}= N_{\text{bulk}} \la{holo}
\eea
where $N_{\text{sur}}$ represents the number of degrees of freedom on the surface with Hubble radius $1/H$,
so $N_{\text{sur}}$  takes the following form

\bea
N_{\text{sur}}= 4 S= \frac{A}{ \ell_{P}^2 }=\frac{4 \pi }{\ell_{P}^2 H^2}\la{Ns}
\eea
where S is the entropy, A is the Hubble area given as $4 \pi H^{-2}$ and $H$ is the Hubble constant.
Note here, we use the entropy area law.
On the other side, the bulk degrees of freedom are given by equipartition law of energy
as follows in the natural units ($k_B=c=\hbar=1$):

\bea
N_{\text{bulk}}=2 ~\frac{ E_{\text{komar}}}{T}
\eea
If the temperature is taken to be Hawking temperature, $T= H/2\pi$ which defines the temperature on the
apparent or Hubble horizon with Komar energy $E_{\text{komar}}= (\rho+3 p) V$, where $V$ is the Hubble volume, and $E_{\text{komar}}$ represents the energy
contained inside this volume $V = 4 \pi / 3 H^3$. For de Sitter case $\rho=-p$, one gets the following relation

\bea
H^2= 8 \pi \ell_{P}^2\rho/3 \la{FRW}
\eea
The last equation represents the Friedmann equations and this shows the consistency
of Holographic principle in Eq. (\ref{holo}) to yield at the end Einstein equation of FRW universe in Eq.(\ref{FRW}).

The validity of holographic equipartition of Eq. (\ref{holo}) for pure de Sitter universe motivates Padmanabhan  to consider
our real universe which is asymptotically de Sitter. The conceptual idea is to assume that
expansion of the universe is equivalent to the emergence of space, and as cosmic time evolves,
the holographic equipartition is obtained at the end. Based on this, it is argued that
dynamical equation which describes the  emergence of space should relate the emergence of space
to the difference between the number of degrees of freedom
in the holographic surface ($N_{\text{sur}}$) and the number of degrees of freedom in the emerged bulk $N_{\text{bulk}}$.
To translate this argument into a dynamical equation,  it is assumed that
during the infinitesimal interval $dt$ of the cosmic time, an increase in the cosmic volume happens to be
\be
\frac{dV}{dt}= \ell_{p}^2 (N_{\text{sur}}-N_{\text{bulk}}) \label{proposal}
\ee
Again, using the expression for cosmic volume $V = 4 \pi / 3 H^3$,
Hawking temperature for Hubble horizon $T= H/2\pi$, number of degrees of freedom on the surface of Eq. (\ref{Ns}) and number of degrees of freedom
in the emerged bulk which is given in terms of Komar energy $N_{\text{bulk}}=2E/T$, where $E= (\rho+3 p) V$, one obtains the following dynamical equation
\bea
\frac{\ddot{a}}{a}= -\frac{4 \pi \ell_{P}^2}{3}(\rho+3p) \la{FRW1}
\eea
The last equation represents the dynamical Friedmann equation of FRW universe. Using continuity equation
\bea
\dot{\rho}=-3 H (\rho+p)\la{FRW2},
\eea
and multiplying Eq. (\ref{FRW1}) with $\dot{a}~ a$, one gets the following equation

\bea
H^2+\frac{k}{a^2}=\frac{8\pi \ell_{P}^2}{3}\rho \label{2FRW}
\eea

In a series of papers the constant $k$ is understood as spatial constant of FRW universe
\cite{Sheykhi:2013tqa,Sheykhi:2013ffa,Ai:2013jha,Sheykhi:2013oac,Eune:2013ima,Ling:2013qoa,Tu:2013gna,Yang:2012wn,Cai:2012ip,Cognola:2013fva}.
We note here that this derivation depends on the fact that number of degrees of freedom on the surface are given by
$N_{\text{sur}}= 4 S= A/4 L_{Pl}^2 =4 \pi/L_{Pl}^2 H^2$. So any kind of corrections
to the entropy-area law should imply corrections to the Friedmann equations. It is worth mentioning that emergence
of cosmic space has been further studied with braneworld scenarios and many other aspects in
\cite{Sheykhi:2013tqa,Sheykhi:2013ffa,Ai:2013jha,Sheykhi:2013oac,Eune:2013ima,Ling:2013qoa,Tu:2013gna,Yang:2012wn,Cai:2012ip,Cognola:2013fva}.
In the next section, we are going to generalize the framework of emergence of cosmic space
for any general form of entropy as a function of area which can host every possible correction to entropy-area law.


\section{General Modified Friedmann equation}
\label{ModFried}

In this section, we study the impact of the most general form of entropy
as a function of area on the emergent cosmic
space and derive the modified dynamical Friedmann equation.
First, let us consider the most general form of entropy-area law. The general form takes the form
 as follows:

\bea
S=\frac{A_{\text{eff}}}{4 \ell_P^2} \label{entropy3}
\eea
where $A_{\text{eff}}$ is a general function of the area $A= 4 \pi r^2=4\pi/H^2$.
In emergence of cosmic space framework, the entropy plays a central role in calculating
the surface degrees of freedom and the area (which is proportional to the entropy)
plays the central role in calculating the bulk degrees of freedom in terms of Komar energy.
Since the surface degrees of freedom is equal to $4 S$ and the bulk degrees of freedom
is equal to Komar energy which is proportional to the cosmic volume and hence the cosmic area.
This tells us that the corrections to degrees of freedom (bulk or surface) should follow
solely from the entropy-area law which may be modified by quantum gravity corrections
as we shall consider in this paper\footnote{We thank the referee for paying our attention for this important note}.

Notice that we assumed the existence of effective area $A_{\text{eff}}$ in Eq. (\ref{entropy3}) instead of normal area $A$.
This definitely assumes an existence of effective cosmic volume $V_{\text{eff}}$ instead of normal cosmic volume $V$. Making use of the definitions
of cosmic (Hubble) area $A= 4 \pi/H^2$ and the cosmic volume $V= 4\pi/(3 H^3)$, we can obtain a
general relation between the Hubble area A and cosmic volume as $V= A^{3/2}/(3 \sqrt{4\pi})$. This relation
could be generalized in the case of effective area and effective cosmic volume to be as follows:

\bea
V_{\text{eff}}= \frac{A_{\text{eff}}^{3/2}}{3 (4\pi)^{1/2}} \label{cvol}
\eea

Now, to establish the emergence of cosmic space in the existence of generic form of entropy as a function
of area, we should have the time evolution of the effective cosmic volume $dV_{\text{eff}}/dt$, the number of degrees
of freedom on the surface $N_{\text{sur}}=4 S$ and finally the number of degrees of freedom
in the bulk $N_{\text{bulk}}$. Let us first calculate the time evolution of effective cosmic volume

\bea
\frac{dV_{\text{eff}}}{dt}=\frac{dV_{\text{eff}}}{dA_{\text{eff}}}~~\frac{dA_{\text{eff}}}{dt}=\frac{1}{2 (4\pi)^{1/2}}A_{\text{eff}}^{1/2} \frac{dA_{\text{eff}}}{dt} \label{Dvol1}
\eea
By looking at Eq. (\ref{Dvol1}) and comparing it with previous studies in \cite{Sheykhi:2013ffa,Ai:2013jha},
we observe that the authors did not consider the most generic form of the time evolution of cosmic volume ($d(V_{\text{eff}})/dt$)
where they used in their derivation the relation $d(V_{\text{eff}})/dA_{\text{eff}}= 1/2H$ which is completely inconsistent,
but as we shown in Eq. (\ref{Dvol1}) that this expression in general takes the form
$d(V_{\text{eff}})/dA_{\text{eff}}=(1/2 (4\pi)^{1/2})A_{\text{eff}}^{1/2}$, and for a special case in which $A_{\text{eff}}=A=4\pi/H^2$,
the changing of cosmic volume reduced to be $d(V_{\text{eff}})/dA_{\text{eff}}= 1/2H$. This of course affects the calculations in \cite{Sheykhi:2013ffa,Ai:2013jha} to be non-exact.

We use Eq. (\ref{entropy3}) to write the last equation in terms of the generic form of the entropy $S$ as follows:

\bea
\frac{dV_{\text{eff}}}{dt}= \frac{4 \ell_P^3}{(4\pi)^{1/2}}\sqrt{S}~~ \frac{dS}{dA}\frac{dA}{dt}\label{Dvol}
\eea
Since we have $A= 4\pi/H^2$, then the time derivative of the area $A$ is given by

\bea
\frac{dA}{dt}= -8 \pi H^{-3} \dot{H}\label{Darea1}
\eea
By substituting Eq. (\ref{Darea1}) into Eq. (\ref{Dvol}), we got the following:

\bea
\frac{dV_{\text{eff}}}{dt}= \frac{4 \ell_P^3}{(4\pi)^{1/2}}\sqrt{S}~~ \frac{dS}{dA}(-8 \pi H^{-3} \dot{H}) \label{Dvolume}
\eea

Now, we calculate the number of degrees of freedom on the surface $N_{\text{sur}}$ in terms of the generic form of the entropy

\bea
N_{\text{sur}}= 4 S= \frac{A_{\text{eff}}}{\ell_{p}^2} \label{surface}
\eea
Turning to calculating the number of degrees of freedom in the bulk for the generic entropy:

\bea
N_{\text{bulk}}= 2\frac{E_{\text{komar}}}{T}
\eea
The most generic form of the Komar energy takes the following form:
\bea
E_{\text{komar}}=-(\rho~+~3p)~V_{\text{eff}} =(\rho~+~3p)~ \frac{A_{\text{eff}}^{3/2}}{3 (4\pi)^{1/2}} \label{komar}
\eea
where we have added minus sign in the Komar energy to have positive $N_{\text{bulk}}$ which makes sense
in an accelerating universe where $\rho+3p < 0$.

The general form of the Hawking temperature  for a generic form of entropy
takes the following expression:

\bea
T= =\frac{\kappa}{8\pi\ell_P^2}\frac{dA}{dS}=\frac{H}{8 \pi \ell_P^2} \frac{dA}{dS} \label{Hawking}
\eea
where $\kappa$ is the surface gravity, and in cosmological apparent horizon it becomes $H$ \cite{Cai}.
So, the generic form of the number of degrees of freedom in the bulk for a generic form of the entropy S
is given by:

\bea
N_{\text{bulk}}= -\frac{16 \pi \ell_P^2}{3H}\frac{A_{\text{eff}}^{3/2}}{(4\pi)^{1/2}} \frac{dS}{dA} (\rho+3p) \label{bulk}
\eea

Using the argument by Padmanabhan \cite{Padmanabhan:2012ik} which proposes that the
dynamical equation which describes the  emergence of space should relate the emergence of space
to the difference between the number of degrees of freedom in the holographic surface
($N_{\text{sur}}$) and the number of degrees of freedom in the emerged bulk $N_{\text{bulk}}$.
We use Eq.(\ref{proposal}) and find that the dynamical equation of FRW universe filled with perfect
fluid with the generic form of entropy is given as follows:
\be
\frac{dV}{dt}= \ell_{p}^2 (N_{sur}-N_{bulk}) \label{proposal1}
\ee
By substituting  Eq. (\ref{Dvolume}), Eq. (\ref{surface}) and Eq. (\ref{bulk}) into Eq. (\ref{proposal1}),
we then get the generic dynamical equation of FRW universe filled with perfect fluid which corresponds to a generic form of
the entropy. This generic dynamical equation takes the following expression:

\bea
&&\frac{4 \ell_P^3}{(4\pi)^{1/2}}\sqrt{S}\frac{dS}{dA}(-8 \pi H^{-3} \dot{H})= \ell_{p}^2 \Bigg(
\frac{A_{\text{eff}}}{\ell_{p}^2} + \nn\\&& \frac{16 \pi \ell_P^2}{3H}\frac{A_{\text{eff}}^{3/2}}{(4\pi)^{1/2}} \frac{dS}{dA} (\rho+3p)\Bigg)
\label{generic}
\eea
If we set $S= \frac{A}{4\ell_{p}^2}$, we get the standard dynamical Friedmann equation

\bea
\frac{\ddot{a}}{a}= -\frac{4 \pi \ell_{P}^2}{3}(\rho+3p) \label{standard}
\eea
So we get the most general dynamical equation of FRW universe filled with perfect fluid
in the framework of emergence of cosmic space. Our general equation in Eq. (\ref{generic})
is different from the result that has been obtained in \cite{Sheykhi:2013ffa} for the following reasons:
\begin{itemize}
\item The authors in \cite{Sheykhi:2013ffa}
did not consider the most generic form of the time evolution of cosmic volume ($d(V_{\text{eff}})/dt$)
where they used in their derivation the relation $d(V_{\text{eff}})/dA_{\text{eff}}= 1/2H$, but as we shown in our paper
in Eq. (\ref{Dvol1}) that this expression in general takes the form
$d(V_{\text{eff}})/dA_{\text{eff}}=\frac{1}{2 (4\pi)^{1/2}}A_{\text{eff}}^{1/2}$, and for a special case in which $A_{\text{eff}}=A=4\pi/H^2$,
the changing of cosmic volume reduced to be $d(V_{\text{eff}})/dA_{\text{eff}}= 1/2H$. This of course affects the calculations in \cite{Sheykhi:2013ffa}
to be non-exact.
\item The other thing which is not exact in \cite{Sheykhi:2013ffa}, the authors calculated Komar energy  in terms of $V$ instead of $V_{\text{eff}}$.
But in our paper, we considered  in Eq. (\ref{komar})  that the Komar energy is defined in terms of the effective volume $V_{\text{eff}}$.
\item The last non-exact thing in \cite{Sheykhi:2013ffa}, the authors considered in their
calculations for $N_{\text{bulk}}$ that the Hawking temperature $T= H/2\pi$ but this is only valid
if we choose only $S=A/4\ell_{p}^2$. This, in fact, is not exact choice,
because the Hawking temperature should follow from the entropy,
and Hawking temperature that corresponds to the general form of the entropy is
given in our paper as we shown in Eq. (\ref{Hawking}).
\end{itemize}
Due to the above reasons, we think that our Eq. (\ref{generic})
is the exact general equation which describes the dynamics of FRW universe filled with perfect fluid for
a general form of the entropy using emergence of cosmic space framework.
We find that Eq. (\ref{generic}) could take a more compact form in terms of a general form of the entropy as
follows:

\bea
\frac{\ell_P}{\sqrt{4\pi}} \frac{dS}{dA}(\frac{-8 \pi\dot{H}}{H^3})= \sqrt{S}+\frac{32~\pi~\ell_{p}^5}{3 H\sqrt{4\pi}}~ S~\frac{dS}{dA}(\rho+3p).~~
\label{compact}
\eea

In the next section, we are going to apply our equation (\ref{generic}) or (\ref{compact}) with the entropy-area relation that follows
from the generalized uncertainty principle (GUP).

\section{Emergence of Cosmic Space and GAUSS-BONNET Gravity}

We investigate in this section the approach that we introduced in the previous section and we derive the modified
Friedmann equation in Gauss-Bonnet gravity. Since the entropy-area law plays the central role in the emergence
of cosmic space framework, we use the one proposed in Gauss-Bonnet gravity for static and spherically symmetric black hole which is given as follows \cite{GaussBonnet}:

\bea
S= \frac{A}{4 \ell_P^{~n-1}}\le(1+\frac{n-1}{n-3} \frac{2\tilde{\a}}{r^2}\ri),
\eea
where $A= n\Omega_n r^{n-1}$ is defined as horizon area and $r$ is called a horizon radius \cite{GaussBonnet},
and the parameter $\tilde{\a}= (n-2)(n-3)\a$ where $\a$ is the Gauss-Bonnet coefficient which takes positive value.
By assuming that the modified entropy-area law due to Gauss-Bonnet gravity would work for the
apparent horizon of FRW universe. Based on this, we may replace the horizon radius $r$ with apparent horizon radius $r_A$.
The entropy-area law  for the apparent horizon will be as follows:
\bea
S= \frac{A}{4 \ell_P^{~n-1}}\le(1+\frac{n-1}{n-3} \frac{2\tilde{\a}}{r_A^2}\ri).
\eea
We use this modified entropy-area law in the emergence of cosmic space framework.
The effective area will be given as follows:
\bea
A_{\text{eff}}= A \le(1+\frac{n-1}{n-3} \frac{2\tilde{\a}}{r_A^2}\ri)= n\Omega_n r^{n-1} \le(1+\frac{n-1}{n-3} \frac{2\tilde{\a}}{r_A^2}\ri).~
\label{effA}
\eea
We use Eq. (\ref{Dvol1}) to calculate the time evolution of effective cosmic volume.
By substituting Eq. (\ref{effA}) into Eq. (\ref{Dvol1}), we get up to the first order of $\a$ the following expression:
\bea
\frac{dV_{\text{eff}}}{dt}&=&\frac{1}{2 (4\pi)^{1/2}}A_{\text{eff}}^{1/2} \frac{dA_{\text{eff}}}{dt}\\&=&
n \Omega_n r_A^{n-1} \dot{r}_A\le[1+\frac{n-2}{n-3} 2 \tilde{\a} r_A^{-2}+O(\tilde{\a}^2)\ri] \label{dvGauss}
\eea
Now, we calculate the modified surface degrees of freedom $N_{\text{sur}}$ using Eq. (\ref{surface}), and we get
\bea
N_{\text{sur}}=4S= \frac{n\Omega_n r_A^{n+1}}{\ell_P^{~n-1}}\le[r_A^{-2}+\frac{n-1}{n-3} 2 \tilde{\a} r_A^{-4}\ri]\label{NGauss}
\eea
The bulk Komar energy in $(n+1)$-dimensions is given by \cite{Nbulk}:
\bea
E_{\text{Komar}}=\frac{(n-2)\rho+n~p}{n-1} V_{\text{eff}},
\eea
where we replaced $V$ with $V_{\text{eff}}$ in the above equation. After few calculations,
it is found the bulk degrees of freedom is given by
\bea
N_{\text{bulk}}&=& 2 \frac{E_{\text{Komar}}}{T}\\&=& -4 \pi \Omega_n r_A^{n+1}\le[1+\frac{n}{n-3}\frac{2\tilde{\a}}{r_A^2}\ri] \nn\\&&\times
\frac{(n-2)\rho+n~p}{n-1}.\label{NbGauss}
\eea
By employing Eqs. (\ref{dvGauss}, \ref{NGauss}, \ref{NbGauss}), in the formula of emergence of cosmic space
that is suggested in \cite{Sheykhi:2013oac}, we get
\bea
\frac{dV_{\text{eff}}}{dt}= \ell_P^{n-1}\frac{r_A}{H^{-1}}\le(N_{\text{sur}}-N_{\text{bulk}}\ri)
\eea
We get the following equation
\bea
&&\dot{r}_A\frac{H^{-1}}{r_A}\le(r_A^{-2}- \frac{2(2 \tilde{\a} r_A^{-4})}{n-3} \ri)-\le(r_A^{-2}-\frac{2 \tilde{\a} r_A^{-4}}{n-3}\ri)+O(\tilde{\a}^2)\nn
\\&&= 4 \pi \ell_P^{n-1}\frac{-2 \rho-\dot{\rho}/H}{n(n-1)}\label{dvdtGauss}
\eea
where we have used the continuity equation in $(n+1)$-dimensions which is given by
\bea
\dot{\rho}+n H (\rho+p)=0
\eea
By multiplying both sides of Eq. (\ref{dvdtGauss}) with the factor $2 \dot{a} a$, and integrate both sides
we get the modified Friedmann equation as follows:
\bea
\frac{d}{dt}\Bigg[a^2\Big(H^2+\frac{k}{a^2}&-& \frac{2 \tilde{\alpha}}{n-3}\le(H^2+\frac{k}{a^2}\ri)^2\Big)\Bigg]+O(\tilde{\alpha}^2)\nn\\&=& \frac{16 \pi \ell_P^{n-1}}{n(n-1)} \frac{d}{dt}(\rho a^2)
\eea
By integrating the above equation, we end with
\bea
\le(H^2+\frac{k}{a^2}\ri)&-& \frac{2}{n-3}\tilde{\alpha}\le(H^2+\frac{k}{a^2}\ri)^2+O(\tilde{\alpha}^2)\nn\\&=& \frac{16 \pi \ell_P^{n-1}}{n(n-1)}\rho
\eea

This equation is similar to the Friedmann equation
in Gauss-Bonnet gravity \cite{Cai} with slight difference in the factor which depends on $1/(n-3)\tilde{\alpha}=(n-2) \alpha$ which
shows that the correction term will be vanishing for $n=2$. This shows that
the approach we are considering may be useful to derive the modified Friedmann equation
in Gauss-Bonnet gravity. However, we should note here that our proposal in Eq. (\ref{Dvol1}) FRW
for Gauss-Bonnet gravity with slight difference in numeric factor from the one derived in \cite{Cai} in
contrast with FRW of Gauss-Bonnet gravity that been derived in \cite{Sheykhi:2013ffa,Sheykhi:2013oac} which agrees in the numeric factor
with the one derived in \cite{Cai}\footnote{We thank the referee for paying our attention to calculate FRW of Gauss-Bonnet gravity which
helped us substantially to write this section }.

\section{Modified Entropy-Area law due to GUP}
\label{GUPentropy}

We review first in this section the generalized uncertainty principle (GUP) \cite{guppapers} and secondly we review
its effect on the area-entropy law \cite{Eliasentropy0,Eliasentropy1,Ali:2012mt0,Ali:2012mt1,Adler,Cavaglia:2003qk}.
We then show a derivation of the  entropy-area law if GUP is taken into consideration
\cite{Eliasentropy0,Eliasentropy1}.  Based on this, we write the exact dynamical equation of FRW universe
if GUP is taken into consideration using Eq. (\ref{compact}).

The GUP is considered as an intriguing prediction of various frame works of quantum gravity
such as string theory and black hole physics \cite{guppapers} leading to
the existence of a minimum measurable length. This in turn leads to a modification of the quantum uncertainty
principle\cite{guppapers,kmm,kempf,brau}:

\bea
\Delta x \gtrsim \frac{\hbar}{\Delta p}\left[1+ \frac{\beta~ \ell_{P}^2}{\hbar^2} (\Delta p)^2\right], \label{GUP}
\eea
where $\ell_{P}$ is the Planck length and $\beta$ is a
dimensionless constant which depends on the quantum gravity theory.
The new correction term in Eq. (\ref{GUP}) turns to be effective when the momentum
and length scales are of order the Planck mass and of the Planck length, respectively.
It was found that Eq.(\ref{GUP}) implies the existence of minimal
length scale as follows:

\bea
\Delta x \gtrsim \Delta x_{min} = 2 \beta~ \ell_{P}
\eea

Recently, a new form of GUP was proposed in
\cite{advplb,Das:2010zf},~which predicts maximum observable momentum, besides the existence of minimal measurable
length, and is consistent with doubly special relativity theories (DSR)\cite{cg}, string theory and black holes physics \cite{guppapers,kmm,kempf,brau}.~It
ensures $[x_i,x_j]=0=[p_i,p_j]$, via the Jacobi identity.
\bea
[x_i, p_j] = i \hbar\hspace{-0.5ex} \left[  \delta_{ij}\hspace{-0.5ex}
- \hspace{-0.5ex} \alpha\hspace{-0.5ex}  \le( p \delta_{ij} +
\frac{p_i p_j}{p} \ri)
+ \alpha^2 \hspace{-0.5ex}
\le( p^2 \delta_{ij}  + 3 p_{i} p_{j} \ri) \hspace{-0.5ex} \ri]~
\label{comm01}
\eea
where $\alpha = {\alpha_0}/{M_{p}c} = {\alpha_0 \ell_{p}}/{\hbar},$
$M_{p}=$ Planck mass, $\ell_{p}=$ Planck length,
and $M_{p} c^2=$ Planck energy.
In a series of papers, various applications of the new model of GUP were investigated \cite{applications}.
For a recent detailed review along the mentioned lines can be found in \cite{amerev}.

The upper bounds on the GUP parameter $\alpha$ has been derived in \cite{Ali:2011fa}.
Moreover, it was investigated that these bounds can
be measured using quantum optics techniques and gravitational wave techniques
in \cite{Pikovski:2011zk,NatureGRW}. This would put several quantum gravity predictions
to test in the laboratory \cite{Pikovski:2011zk,NatureGRW}. Definitely,
this is considered as a milestone in the road of quantum gravity phenomenology.

It has been found in \cite{Ali:2012mt0,Ali:2012mt1}, that the inequality which
would correspond to Eq. (\ref{comm01}) can be written as follows:

\bea \D x \D p \geq \frac{\hbar}{2} \Big[1&&- \a_0 ~\ell_P~
\le(\frac{4}{3}\ri)~\sqrt{\mu}~~ \frac{\D p}{\hbar}\nn\\&&+ ~2~
(1+\mu)~ \a_0^2 ~\ell_P^2 ~ \frac{\D p^2}{\hbar^2} \Big]\,.
\la{ineqII} \eea The last inequality is ( and as far as we know
the only one) following from Eq.\ (\ref{comm01}). The parameter $\mu$
is defined in \cite{Ali:2012mt0}.
Solving it  as a quadratic equation in $\D p$ results in
\bea \frac{\D p}{\hbar}\geq&&\frac{2 \D
x+\a_0
~\ell_P~\le(\frac{4}{3}~\sqrt{\mu}~\ri)}{4~(1+\mu)~\a_0^2~\ell_{p}^2}\times \nn\\ &&\le(1-
\sqrt{1-\frac{8~(1+\mu)~\a_0^2\ell_{p}^2} {\le(2 \D x+\a_0
\ell_P\le(\frac{4}{3}\ri) ~\sqrt{\mu}~\ri)^2}}\ri)
\,.\la{gupso} \eea
The negatively-signed solution is considered as the one that refers to the standard uncertainty relation as $\ell_P/\D x \rightarrow 0$. Using the Taylor expansion, we obviously find that
\be
\Delta p \geq \frac{1}{\Delta x} \le(1- \frac{2}{3}\a_0 \ell_P \sqrt{\mu} \frac{1}{\Delta x} \ri).
\ee
%
%
%
%
%
%
%
%
%
%
%
There have been much studies devoted to study the impact of GUP on the black hole thermodynamics and
to the Bekenstein--Hawking (black hole) entropy (e.g., \cite{Eliasentropy0,Eliasentropy1,Adler,Cavaglia:2003qk,Ali:2012mt0,Ali:2012mt1}).
These studies are based on the argument that Hawking radiation is a quantum process
and it should respect the uncertainty principle.
According to \cite{Eliasentropy0,Eliasentropy1}, a photon is used to ascertain the position of a quantum particle of energy $E$
and according to the argument in \cite{AmelinoCamelia:2004xx} which states that
the uncertainty principle $\D p \geq 1/ \D x$  can be translated to the lower
bound $E\geq 1/ \D x$, one can write for the GUP case:

\bea
E \gtrsim \frac{1}{\Delta x} \le(1- \frac{2}{3}\a_0 \ell_P \sqrt{\mu} \frac{1}{\Delta x} \ri) \label{energy0}
\eea
During absorption process of quantum particle with energy $E$ and size $R$ by the black hole, it supposed for black hole area to increase
by the following amount

\bea
\Delta A \geq 8 \pi\, \ell_P^2\, E\, R,
\eea
The quantum particle itself implies the existence of finite bound given by
\be
\Delta A_{min} \geq 8 \pi\, \ell_P^2\, E\, \Delta\, x. \la{Darea}
\ee
Using the Eq. (\ref{energy0}) in the inequality (\ref{Darea}), we get
\bea
\Delta A_{min} \gtrsim 8 \pi\, \ell_P^2 \le(1- \frac{2}{3}\a_0 \ell_P \sqrt{\mu} \frac{1}{\Delta x} \ri)   \label{Area}
\eea
The value of $\Delta x$ is taken to be inverse of surface gravity $\D x= \kappa^{-1}= 2 r_s$ where $r_s$ is the
Schwarzschild radius, where this is  probably the most sensible choice of length scale
in the context of near-horizon geometry \cite{Eliasentropy0,Eliasentropy1,Adler,Cavaglia:2003qk,Ali:2012mt0,Ali:2012mt1}.
This implies the following
\be
\D x^2 = \frac{A}{\pi} \label{DX}
\ee
Substituting  Eq. (\ref{DX}) into Eq. (\ref{Area}), we get

\be
\Delta A_{min} \simeq \lambda \ell_{p}^2 \le[1- \frac{2}{3}\, \a_0\, \ell_P\, \sqrt{\frac{ \mu \, \pi}{A}}\ri],
\ee
%
%
where the parameter $\lambda$ will be fixed later from the Bekenstein-Hawking entropy formula.
According to \cite{Bekenstein,Hawking}, the black hole's entropy is conjectured to depend on the
horizon's area. From the information theory \cite{Adami:2004mx}, it has been found
that the minimal increase of entropy should be independent on the area. It is just
one "bit" of information which is $\Delta S_{min}=b = \ln(2)$.

\be
\frac{dS}{dA}= \frac{\Delta S_{min}}{\Delta A_{min}} = \frac{b}
{ \lambda \ell_{p}^2 \le[1- \frac{2}{3}\, \a_0\, \ell_P\, \sqrt{\frac{ \mu \, \pi}{A}}\ri]}.
\ee
According to \cite{Eliasentropy0}, the Bekenstein-Hawking entropy formula has been used
to calibrate the constants $b/\lambda= 4$, so we have
\bea
\frac{dS}{dA}= \frac{\Delta S_{min}}{\Delta A_{min}} = \frac{1}
{4 \ell_{p}^2 \le[1- \frac{2}{3}\, \a_0\, \ell_P\, \sqrt{\frac{ \mu \, \pi}{A}}\ri]} \label{Dentropy}
\eea
%
In this paper, we are interested with the modified Bekenstein-Hawking entropy law due to GUP and its impact on
dynamical Friedmann equation in emergence of cosmic space framework. By  integrating Eq. (\ref{Dentropy}),
to yield the modified Bekenstein-Hawking entropy law
due to GUP up to the first order of $\a$, we get
\be
S=\frac{A}{4\, \ell_P^2} + \frac{2}{3}\, \a_0\,  \sqrt{\pi\, \mu\, \frac{A}{4\, \ell_P^2}}. \la{correctENTROPY}
\ee
We find that the entropy is directly related to the area and gets a correction when applying GUP-approach.

We note here  the considered model of GUP introduces corrections to the entropy-area law proportional to first order of Planck length as in Eq. (\ref{correctENTROPY}) in our revised version of the paper. This equation says that GUP introduces a correction at the first order of Planck length, where Eq. (\ref{correctENTROPY}) can be written as follows:
\bea
S= \frac{A}{4\, \ell_P^2} \le[1+ \frac{2}{3} \alpha_0 \sqrt{\pi\, \mu}~  \ell_P~  A^{-1/2}\ri]
\eea
The other well known corrections to the entropy-area law are known as logarithmic corrections and they arise from the loop quantum gravity due to thermal equilibrium fluctuations and quantum fluctuations\cite{log}. The logarithmic correction  plays its role starting from the second order of Planck length as it is shown in the following equation \cite{Sheykhi:2013ffa,log}
\bea
S= \frac{A}{4\, \ell_P^2} \le[1+ c~ \frac{4 }{A} ~\ell_P^2~ \ln{\frac{A}{4\, \ell_P^2}}+ d~~\ell_P^4~ \frac{4 }{A^2} \ \ri]
\eea
So they are not relevant to study them at the same order with the GUP,
where GUP plays its role at the first order of Planck scale but the logarithmic
corrections play their role starting from the second order of Planck length.
This means that the GUP corrections could be reliable up to the first order
of Planck length. For more details on the logarithmic correction and the numeric values
of dimensionless constants $c$ and $d$, this Ref. \cite{Cai:2008ys} may be consulted.

In the next section, we implement Eq. (\ref{correctENTROPY}) in our derived Eq. (\ref{compact})
to derive the modified Friedmann  equation due to GUP, then we derive the corresponding Raychaudhuri equation
to study whether the FRW universe has a singularity or not using the fixed point method \cite{Awad:2013tha}.

\section{Modified Friedmann equation due to GUP}
\label{FRW-GUP}

In this section, we are going to implement the modified Bekenstein-Hawking entropy area law of Eq. (\ref{correctENTROPY})
with the general dynamical Friedmann equation that we derived in Eq. (\ref{compact}). First,
we set $\gamma= 2/3 \a_0 \sqrt{\pi \mu}$. After few calculations and using Eqs. (\ref{Dentropy}) and (\ref{correctENTROPY}), the modified Freidmann equation due to GUP will be as follows:

\bea
\frac{\ddot a}{a}= \dot{H}+H^2= -\frac{4 \pi \ell_P^2}{3} \le(\rho+3p\ri) \le(1+\frac{2\gamma \ell_P}{\sqrt{4\pi}}H\ri) \label{GUPFRW1}
\eea

To find the modified Friedmann equation which is analogue to Eq. (\ref{2FRW}), we just multiply Eq. (\ref{GUPFRW1})
by $a\dot{a}$ and then integrate the equation. We get the following:

\bea
\frac{d}{dt}(\dot{a}^2)\le(1-\frac{2\gamma \ell_P}{\sqrt{4\pi}}\frac{\dot{a}}{a}\ri)=\frac{8 \pi \ell_P^2}{3}\frac{d}{dt}(\rho a^2)
\eea
By integrating the last equation, we get:
\bea
&&\dot{a}^2\le(1-\frac{2\gamma \ell_P}{\dot{a}^2 a\sqrt{4\pi}}\int \dot{a} d(\dot{a}^2)\ri)= \frac{8 \pi \ell_P^2}{3}\rho\\
&&H^2 (1-\frac{4\gamma \ell_P}{3\sqrt{4\pi}}H)+\frac{k}{a^2}= \frac{8 \pi \ell_P^2}{3}\rho \label{GUPFRW2}
\eea

To study the applicability of the modified Friedmann equations that we obtained in Eqs. (\ref{GUPFRW1}) and (\ref{GUPFRW2}), we derive the
Raychaudhuri equation that corresponds to the modified Friedmann equations and study the solutions if they have singularities.

It is constructive to discuss the general conditions that lead to a nonsingular cosmology.
The general Raychaudhuri equation for a general form of the entropy in emergence of cosmic space framework
could take the following form

\bea
\dot{H}=-F(H) , \label{Conserv11}
\eea
where $F(H)$ for standard Friedmann equation takes the form $F(H)=-3/2 (1+\omega) H^2$ for equation
of state $\rho=\omega p$ and this definitely has a singularity. For a general $F(H)$, it introduces first-order system which is well studied in dynamical system (see  e.g., \cite{Awad:2013tha}, or see \cite{strogatz} for more general applications) in cosmological contexts. Knowing the fixed points of the function $F(H)$, (i.e., its zeros, let us call them $H_i$) and its asymptotic behavior enables one to qualitatively describe the behavior of the general solution without actually solving the system. Fixed points are classified according to their stability to stable, unstable, or half-stable. In \cite{Awad:2013tha} a very similar system has been studied which was expressed in terms of the Hubble rate. It is straightforward to use the same analysis to study the density $\rho$ instead of the Hubble rate $H$.

Our basic idea for resolving finite-time singularities is to show the existence of an upper bound for the density $H$ (through having a fixed point $H_1$) which is reached at an infinite time, or to show the existence of a point at which the density is unbounded (a potential singularity) but reached in an infinite time, i.e., not a physical singularity. Therefore, following the discussion in \cite{Awad:2013tha}, one can show that finite-time singularities are absent if $F(H)$ has a fixed points that can be reached in infinite time.

Turning into our modified Friedmann equations of Eqs. (\ref{GUPFRW1}) and (\ref{GUPFRW2}), and with
considering equation of state of the perfect fluid as $\rho=\omega p$ and by setting the constant
$k=0$ , the corresponding
Raychaudhuri equation will be

\bea
\dot{H}= -\frac{3}{2} \le(1+\omega\ri)H^2\le(1+\frac{2\gamma \ell_P (1+3\omega)}{9 \sqrt{4\pi}(1+\omega)}H\ri) \label{Raychd}
\eea

\begin{figure}
\includegraphics[scale=0.35,angle=0]{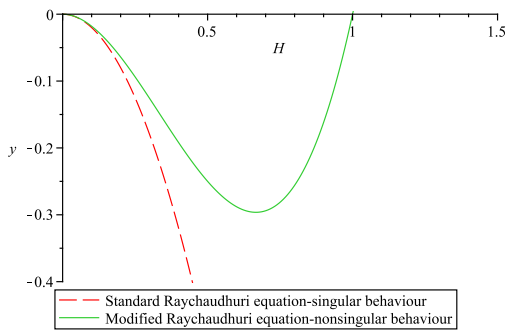}
\caption{$\dot{H}$ versus $H$}.
\label{solution}
\end{figure}

One can observe that the above Raychaudhuri equation (\ref{Raychd}) might be able to resolve the FRW singularities
since it has two fixed points. To see that let us first plot $\dot{H}$ versus $H$ in Fig. \ref{solution}, where we consider the  case $\omega=-2/3$ and $2\gamma \ell_P/\sqrt{4\pi}=1$. From the plot or simple analysis one can observe that the Hubble parameter has a maximum bound which
introduces a cutoff proportional to the GUP parameter $\alpha$.

One can observe that the above system has two fixed points, $H_1=0$ and $H_2\sim 1/ \gamma \ell_P=H_P$, which is showing that the solution is nonsingular and interpolate between $H=0$ and $H_P$. Let us consider the  case $\omega=-2/3$, where the relation between $\dot{H}$ and $H$ is depicted in Fig. (\ref{solution}). In fact this behavior is the same for values of $\omega$ between $-1<\omega<-1/3$ which introduce a range of nonsingular solutions. One can show the absence of finite-time singularities by calculating the time necessary to reach any of the two fixed point $H_f=0$ or $H_P= 1/(\gamma \ell_P)$ (starting from any finite value of Hubble parameter $H^{\star}$)

\be
t= -\frac{2}{3(1+\omega)} \int_{H^{\star}}^{H_f} \frac{dH}{H^2\le(1+\frac{2\gamma \ell_P (1+3\omega)}{9 \sqrt{4\pi}(1+\omega)}H\ri)} =\infty
\ee
which means that the time necessary to reach a fixed point is infinite. This introduces
a possible resolution for  singularity in FRW universe for specific range of $\omega$. This gives solutions which are non-singular and have two fixed points.
We got a similar nonsingular solutions \cite{Awad:2013nxa} in a different framework which is called gravity rainbow \cite{Magueijo:2002xx}.
Also similar nonsingular behavior can be obtained in the framework of nonsingular viscous fluids in Cosmology in references\cite{Awad:2013tha,nsviscous}.

\section{Conclusions}

In this paper, we tackle the idea of generalizing the framework of emergence of cosmic space
for a general form of the entropy as a function of area. We got an exact and general dynamical equation
of FRW universe filled with a perfect fluid and we compared our general equation
with the previous studies in \cite{Sheykhi:2013ffa,Ai:2013jha}.
We investigated the  Einstein-Gauss-Bonnet (EGB)theory which gives a correction to
the entropy-area law by a term which is proportional to $A^2$ as indicated in ref \cite{Cai:2012ip}, and calculated
the modified Friedmann equation in Gauss-Bonnet gravity. We derived a modified Friedmann equation similar to the corresponding Friedmann
equation in Gauss-Bonnet gravity that has been derived in \cite{Cai,Sheykhi:2013ffa,Sheykhi:2013oac} with slight difference in the numeric factor in front of $(H^2+k/a^2)^2$-term.

We then apply this general equation with the corrected entropy-area law due to GUP.
We note that the derived correction terms for Friedmann equation vanishes rapidly with increasing of the apparent radius $(r=1/H)$, as expected. This means that the corrections become relevant at the early universe, in particular, with the inflationary models where the physical scales are few ordered of magnitude less than the Planck scale. When the universe becomes large, these corrections can be ignored and the modified Friedmann equation reduces to the standard Friedmann equation. We can understand that when $a(t)$ is large, it is difficult to excite these modes and hence, the low-energy modes dominate the entropy. But at the early universe, a large number of excited modes can contribute causing a modification to the area law \cite{Sheykhi:2010wm,Sheykhi:2010yq} and hence modified Friedmann equations according to the emergence of cosmic space framework. But could we observe the impact of these corrections on the early universe.  Since these corrections modify the standard FRW cosmology, especially in early times, it is expected to have some consequences on inflation. One of the interesting results reported in the Planck 2013 \cite{Ade:2013uln} is that exact scale-invariance of the scalar power spectrum has been ruled out by more than $5\sigma$. Meaning that, the early universe tiny quantum fluctuations, which eventually cause the formation of galaxies, not only depend on the mode wave number k, but also on some physical scale! This shows that scalar power spectrum and other inflation parameters could depend on physical scale. The energy scale of inflation models has to be around Grand Unified Theories (GUT) scale or larger, therefore, this cutoff scale could be the Planck scale. This indicates that GUP could be an important to be studied  with Friedmann equation as a quantum correction where GUP introduces an existence of minimal length scale which may be the Planck length.

When studying the modified Friedmann equation due to GUP, we got non-singular solutions for a range of values for the equation of state parameter $-1<\omega< -1/3$. Using the analysis in \cite{Awad:2013tha} we find the system exhibits two fixed points, one of them is around the GUP parameter (i.e. Planck scale). Also the system takes infinite time to reach the fixed points which represents a non-singular solution. So we find a possible resolution of FRW singularities due to the effect of GUP.

It would be appropriate to apply our general dynamical equation of FRW universe
in cases of the quantum corrections to the entropy-area law
such as logarithmic corrections and power-law corrections which follows from string theory and loop
quantum gravity, etc. We hope to report on these in the future.

\no {\bf Acknowledgments}\\
The author gratefully thanks the anonymous referee for useful comments and suggestions
which substantially helped in proving the quality of the paper.
The author gratefully thanks Adel Awad for many enlightening
discussions on the subject. This work is supported by Zewail City of Science and
Technology and by Benha University(www.bu.edu.eg), Egypt.



\end{document}